\documentclass[fleqn,twoside]{article}
\usepackage{amsmath}
\usepackage{espcrc2}

\usepackage{epsf}
\usepackage{epsfig,graphics}

\newcommand{\be}{\begin{equation}}
\newcommand{\ee}{\end{equation}}

\renewcommand{\>}{\rangle}

\title{Electromagnetic Hadronic Form-Factors}

\author{Robert G. Edwards\address{Thomas Jefferson National Accelerator
                                  Facility, Newport News, VA 23606, USA},
for the Lattice Hadron Physics Collaboration
}

\begin{document}

\begin{abstract}
We present a calculation of the nucleon electromagnetic form-factors
as well as the pion and rho to pion transition form-factors 
in a hybrid calculation with domain wall valence quarks and improved
staggered (Asqtad) sea quarks.
\end{abstract}

\maketitle

The determination of light hadron physics properties is of considerable
interest to experimental labs such as Jefferson Lab. Many quantities,
such as hadron form-factors and generalized structure functions,
are not well understood theoretically. 
The available and
forthcoming high precision experimental data presents considerable challenges 
and many opportunities for corresponding lattice calculations. 
Including the contribution of light dynamical quarks is an important goal 
of this work, but the cost of generating these gauge ensembles is large.
We have adopted a so-called ``hybrid'' scheme where staggered sea quarks
are used (Asqtad action~\cite{Bernard:2001av}) and domain wall valence 
quarks~\cite{Negele:2004iu}. 
While unitarity is broken at finite lattice spacing, it is recovered in
the continuum limit when the sea and valence quark masses are properly tuned.

In this contribution, we study the efficacy of an uncommon method of
calculating three-point functions that avoids sequential source
techniques. We calculate various electro-magnetic
form-factors without the cost of a new sequential
inversion for each new set of observables. The method is particularly
suitable for valence quark actions amenable to a multi-mass solver,
such as the Overlap quark action. The conclusion is that the method appears 
viable, but very light quark mass tests are needed (and on-going).

A typical sequential source method for computing a matrix element such 
as $\<B(p_f)|V(q)|A(p_i)\>$ for an initial state $A$ with three-momenta 
$p_i$ and final state $B$ at momenta $p_f$ with a two-quark insertion $V$
involves either a sequential inversion through the insertion or the sink.
To properly extract the matrix element for a set of momenta $q=p_f-p_i$, usually
the sequential source for the sequential insertion is held at a fixed momenta.
There are two common techniques. Sequential inversion through insertion: 
benefits are that one can vary the source and sink fields but the 
insertion momenta and operator are fixed. Sequential inversion through sink:
benefits are one can vary the insertion operator and momenta, but the
sink operator and momenta are fixed. In addition, for baryons the
spin projection matrix between the source and sink must be fixed necessitating
different sequential inversions to extract electric and magnetic quantities.
The common problem is one vertex must have a definite momentum.

As an alternative, one can instead make a sink (or source) quark {\em propagator}
(as opposed to the state) have a definite momentum. Putting the sink (or source) 
quarks at definite momentum implies the hadron state has a
fixed momenta via the translation/momentum operator on the lattice. The scheme then is
to build the desired hadron state propagating from the source and sink
where the quark propagators used for one of these states has a fixed momenta.
Thus, one avoids sequential inversions computing $\<B(p_f)|V(q)|A(p_i)\>$. For
example, fixing the sink momenta to a fixed $p_f$, one can extract the
matrix element at all $q$ by vary $p_i$ according to momentum conservation.

To fix a quark's momenta necessarily implies a wall source thus requiring
the need for gauge fixing. Additional tricks can be used
to improve statistics like using charge conjugation and time reversal  (CT)
in (anti-)periodic boundary conditions. This method also works for
Dirichlet boundary conditions where one maintains equal source and sink
separation from the Dirichlet wall. E.g., a wall sink
hadron because a wall source after CT.
If applicable, multi-mass inversion techniques can be applied both
to the source and the sink propagator calculations.

\begin{figure}[t]
\epsfig{file=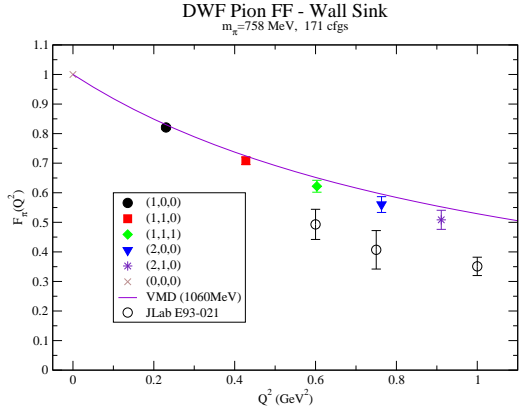, clip=, angle=270, width=8cm}
\vskip -8mm
\caption{Pion electromagnetic form-factor
computed by Eq.~\ref{eq:ratio} and imposing the lattice dispersion
relation.}
\label{fig:pion}
\end{figure}

\begin{figure}[t]
\epsfig{file=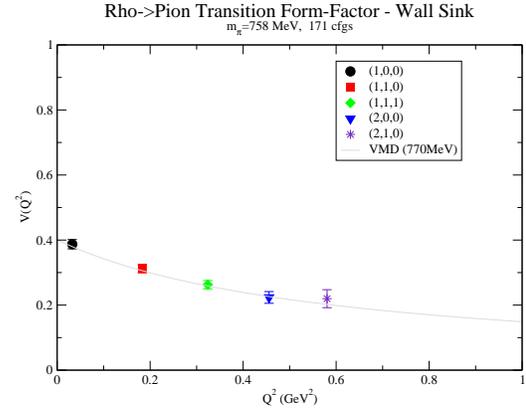, clip=, angle=270, width=8cm}
\vskip -8mm
\caption{$\rho\rightarrow\pi$ transition form-factor. The VMD line
has a 0.4 normalization.}
\label{fig:rhopi}
\end{figure}

For our calculation of hadronic form factors with unquenched gauge
configurations, we performed a hybrid calculation using MILC
$N_f=2+1$ configurations in 20$^3\times$64 volumes, generated
with staggered \texttt{asqtad} sea quarks\cite{Bernard:2001av},
and domain wall valence quarks with domain wall height $m_0=1.7$
and extent $L_s = 16$ of the extra dimension \cite{Negele:2004iu}.
The MILC configurations were HYP blocked \cite{Hasenfratz:2001hp}
before valence propagators were computed, otherwise the residual
chiral symmetry breaking would have been unacceptably large.
Dirichlet boundary conditions were imposed 32 time slices apart. The
lattice spacing scale is set at $a=0.125$fm. The Asqtad set has $m_u=0.01$ 
and $m_s=0.05$, $\beta=6.76$. The valence pion and rho mass are
$m_\pi=758(5)$MeV and $m_\rho=1060(14)$MeV, resp.

The technique for extracting form factors uses
the ratio method of correlation functions:
\be
R^{XY}_{\alpha\beta\mu} = 
  \frac{\Gamma_{\alpha\beta,\mu,AB}^{XY}(t_i,t,t_f,\vec{p}_i,\vec{p}_f)
  \Gamma_{CL}^{YY}(t_i,t,\vec{p}_f)}{
  \Gamma_{AL}^{XX}(t_i,t,\vec{p}_i) \Gamma_{CB}^{YY}(t_i,t_f,\vec{p}_f)}
\label{eq:ratio}
\ee
where $A$, $B$, $C$ are generic smearing labels, $L$ is local and $\mu$ is the direction
of the local current, the initial and final states are $X_\alpha$ and $Y_\beta$. 
Similarly for $R^{YX} = R^{XY}$ where $Y\rightarrow X$ . 
Note, momenta and smearing labels are not interchanged.
In the case of $Y=X$, this ratio cancels all wave-function factors
and exponentials. For the transition case, the
combination  $(R^{YX} R^{XY})^{1/2}$  cancels all 
exponentials and wave-function factors taking into the effects of smearing
on the wave-functions. One extracts the form-factors up to some kinematic factors
using properties of the spin projection matrix between $X_\alpha$ and $Y_\beta$.
In all the subsequent work, we fixed the wall sink to have $\vec{p}_f=0$. 
The source quarks are APE smeared.

The first test of the wall-sink method involves the pion form factor.
The result shown in Fig.~\ref{fig:pion} compares favorably with
a sequential-sink technique~\cite{Bonnet:2003pf} at the
one mass in common corresponding to $am_{\rm val}=0.081$. The
obtained value of $Z_V$ for our local current also agrees well.

\begin{figure}
\epsfig{file=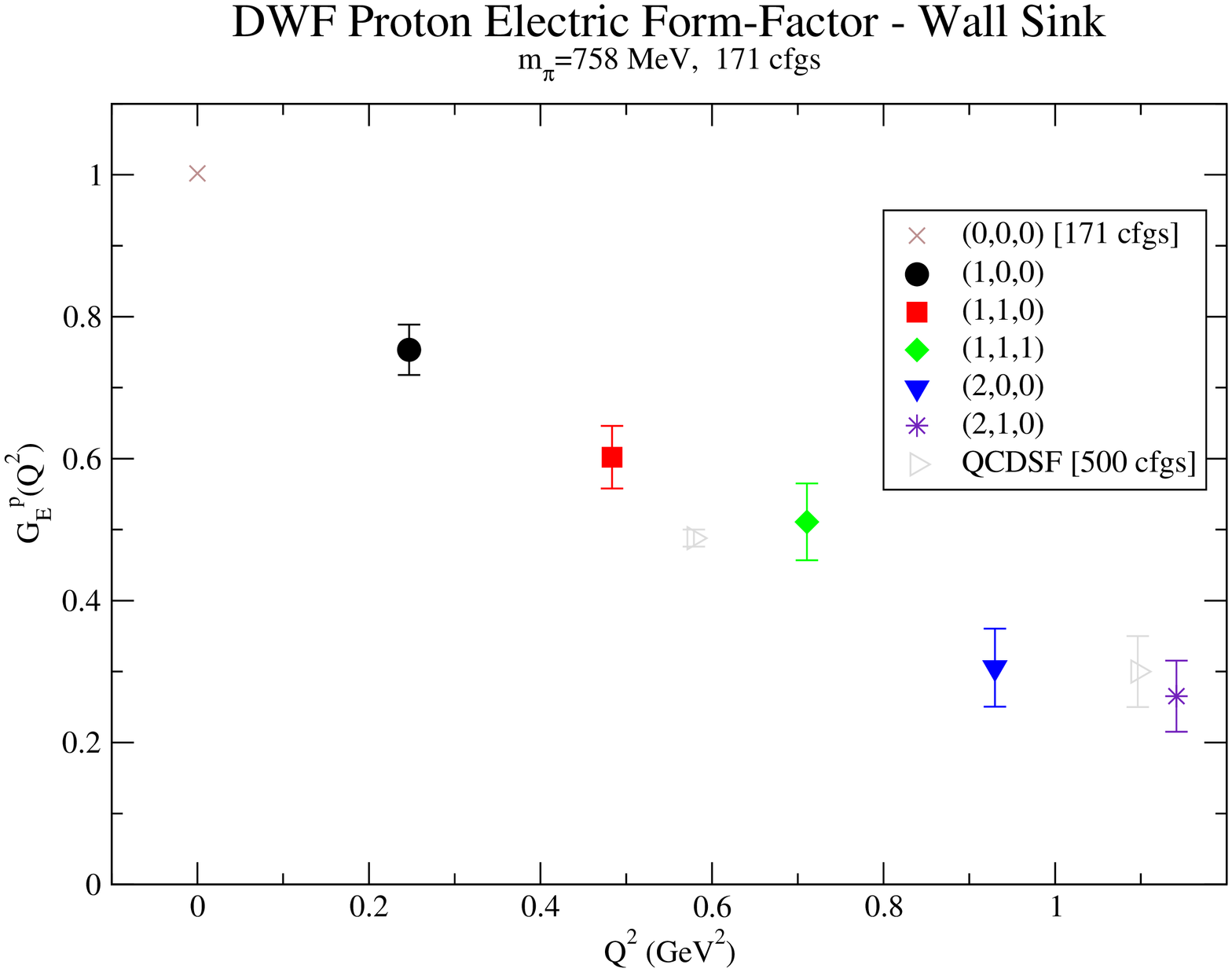, clip=, angle=270, width=8cm}
\vskip -8mm
\epsfig{file=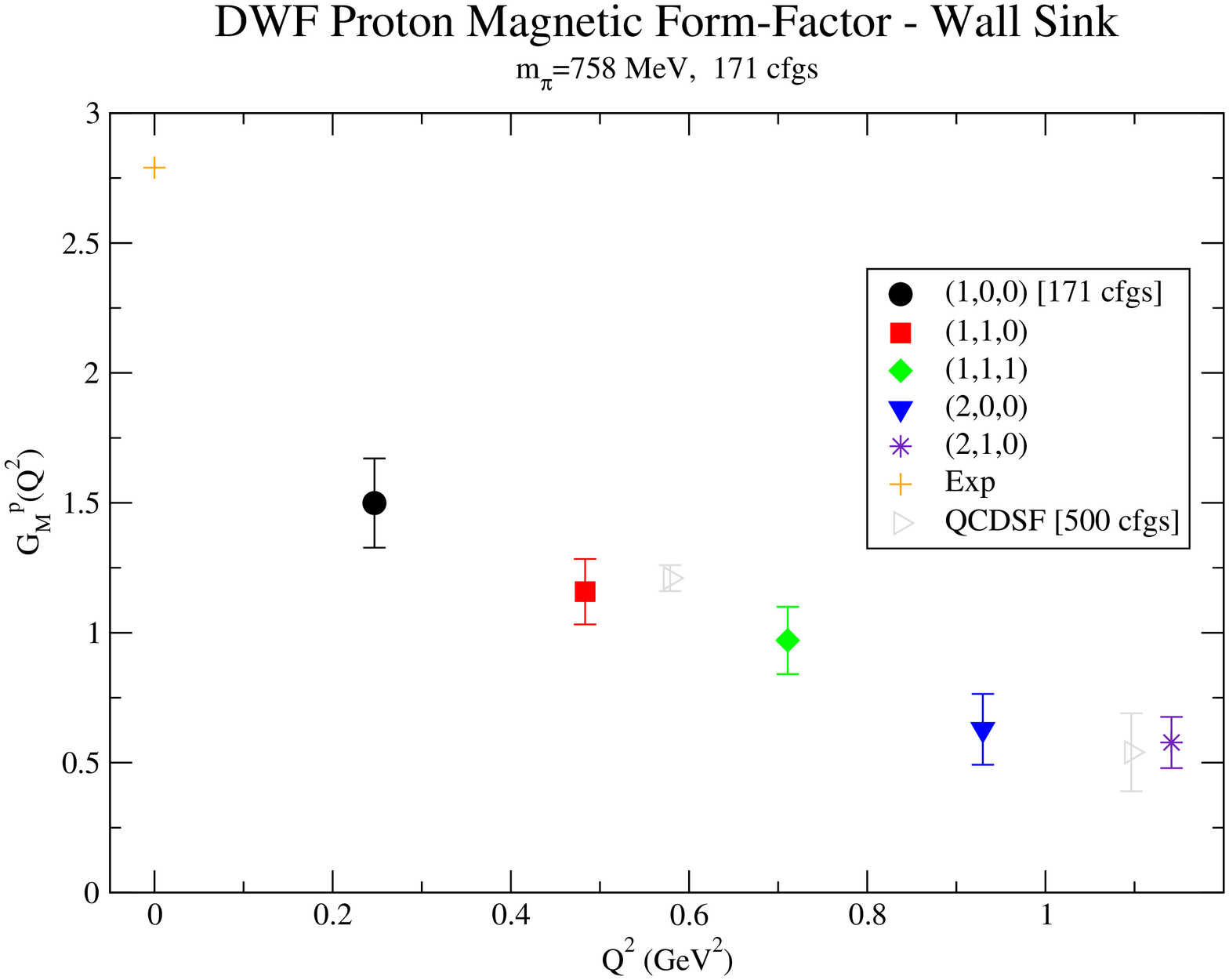, clip=, angle=270, width=8cm}
\vskip -8mm
\caption{Proton electric and magnetic form factors. Also shown
are results from QCDSF~\cite{QCDSF:2003m}.}
\label{fig:proton}
\end{figure}

\begin{figure}
\epsfig{file=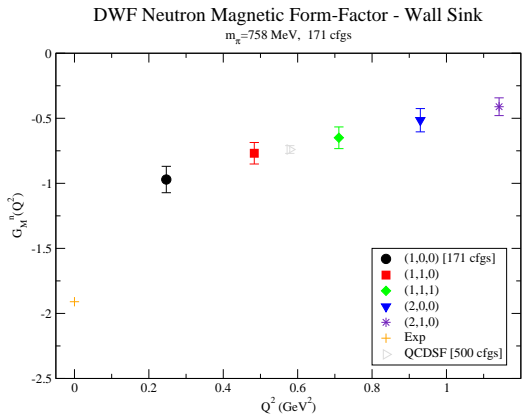, clip=, angle=270, width=8cm}
\vskip -8mm
\caption{Neutron  magnetic form-factor.}
\label{fig:neutron}
\end{figure}

Electro-disintegration of the deuteron has been intensively studied
experimentally. While isovector meson exchange currents
have been identified in such systems, the role of isoscalar meson exchange
is not clear. Here, we provide a lattice measurement of the 
$\rho\rightarrow\pi$ transition form factor which is an important step
in an analysis of exchange mechanisms. The matrix element computed is
\begin{multline*}
Z_V\left<\pi(\vec{p}_f)\left|V_\mu(0)\right|\rho(\vec{p}_i,r)\right> \\
= \frac{2 V(Q^2)}{m_\pi+m_\rho}\epsilon^{\mu\alpha\delta\beta}p_{\alpha,f} 
  p_{\delta,i}\epsilon_\beta(\vec{p}_i,r)
\end{multline*}
The form factor $V(Q^2)$ shown in Fig.~\ref{fig:rhopi} is in
rough agreement with a phenomenological calculation in Ref.~\cite{Gross:1993}.
The coupling constant normalization taken from $V(0)$ is in rough
agreement with experiment Ref.~\cite{Berg:1980,Gross:1993}. 
Higher $Q^2$ are needed.

The extraction of the proton and neutron form factors uses
Eq.~\ref{eq:ratio} together with a spin projection matrix
$T_{\alpha\beta}$ between source and sink nucleon states $N_\alpha$
and $N_\beta$. We can increase statistics on
the electric and magnetic form-factors by averaging over all the
nucleon spin polarizations. The results for the proton and neutron are
shown in Figs.~\ref{fig:proton} and \ref{fig:neutron}. Compared
with quenched clover results from QCDSF~\cite{QCDSF:2003m} 
at roughly the same pion mass ($\beta=6.0$, $\kappa=0.1338$), 
the error bars stay somewhat constant at higher $Q^2$. 
Reasonable agreement is seen with experiment and Ref.~\cite{QCDSF:2003m}.
In addition, a signal is seen for the computationally difficult 
neutron electric form-factor.

This work was supported by the U.S. Department of Energy under
contract DE-AC05-84ER40150. Computations were performed on the 
Pentium IV clusters at JLab.

\end{document}